\documentstyle[12pt]{article}
\begin{document}
\newcommand{\ECM}{\em Departament d'Estructura i Constituents de la
Mat\`eria
                  \\ Facultat de F\'\i sica, Universitat de Barcelona \\
                     Diagonal 647, E-08028 Barcelona, Spain \\
                                      and           \\
                                    I. F. A. E.}

\def\thefootnote{\fnsymbol{footnote}}
\pagestyle{empty}
{\hfill \parbox{6cm}{\begin{center} July 2000
                     \end{center}}}
\vspace{1.5cm}

\begin{center}
\large{Triviality of GHZ operators of higher spin}

\vskip .6truein
\centerline {J. Savinien, J. Taron{\footnote{e-mail: taron@ecm.ub.es}},
             R. Tarrach}
\end{center}
\vspace{.3cm}
\begin{center}
\ECM
\end{center}
\vspace{1.5cm}

\centerline{\bf Abstract}
\medskip

We prove that local observables of
the set of GHZ operators for particles of spin higher than
1/2 reduce to direct sums of the spin 1/2 operators $\sigma_x$, $\sigma_y$
and, therefore, no new contradictions with local realism
arise by considering them.

\vspace{2cm}

\newpage
\pagestyle{plain}

\section{Introduction}
The GHZ theorem \cite{ghz} provides a powerful test of quantum non-locality,
which can be confirmed or
refuted by the outcome of just one single experiment \cite{mermin}.
Formulated for three
spin 1/2 particles \cite{mermin} \cite{peres},
the argument is based on the anti-commutative nature of
the 2x2 spin operators $\sigma_x$, $\sigma_y$. The values of the
three mutually commuting observables
\begin{equation}
\sigma_x^a \otimes \sigma_y^b \otimes \sigma_y^c \equiv
\sigma_x^a \sigma_y^b \sigma_y^c, \;\;\;
\sigma_y^a \sigma_x^b \sigma_y^c, \;\;\;
\sigma_y^a \sigma_y^b \sigma_x^c,
\label{tres}
\end{equation}
and their product, $-\sigma_x^a \sigma_x^b \sigma_x^c$, 
cannot be obtained, consistently, by making 
local assignments to each of the individual spin operators, $m_x^I, \;
m_y^I=\pm 1$, $I=a,b,c$.
This is not
a contradiction of Quantum Mechanics: the state $|\psi\rangle=
\frac{1}{\sqrt{2}} \left( |\uparrow \uparrow \uparrow \rangle -
| \downarrow \downarrow \downarrow \rangle \right)$, for instance,
is one of the common
eigenstates of the four operators, with eigenvalues $\lambda_1=\lambda_2=
\lambda_3=1$, $\lambda_4=-1$, respectively. $|\psi \rangle$ is a highly
correlated (entangled) state of the three parties which has no
defined value for $\sigma_x^I, \sigma_y^I$.

In this note we address the question of how to generalize the argument to
particles of higher spin and find that there are no non-trivial extensions
other than direct sums of operators that can be brought into the form 
$\sigma_x$,$\sigma_y$ by means of local unitarity transformations.
(For odd dimensional Hilbert spaces the direct sum is completed by a 
one-dimensional submatrix, i.e., a c-number in the diagonal).
We give a proof for the cases of spin 1 and 3/2. Similar problems have been
addressed in \cite{adan}.

Let us look for observables $A$, $B$ such that $AB=\omega BA$ 
(their hermiticity
implies that $\omega$ is at most a phase): this is a necessary
condition for the commutator relations 
$[A_1^a A_2^b A_3^c,B_1^a B_2^b B_3^c]= {\rm etc}...=0$ to hold. As
we shall see,
all interesting cases correspond to $\omega=-1$. 
Without loss of generality, $A$ can always be taken diagonal, 
$A={\rm diag}(\lambda_1,\lambda_2)$,
for the simplest case s=1/2. The above condition reads
\begin{equation}
AB-\omega BA=\left(
\begin{array}{cc}
(1-\omega) \lambda_1 b_{11} & (\lambda_1-\omega \lambda_2) b_{12} \\
(\lambda_2-\omega \lambda_1) b_{12}^* & (1-\omega) \lambda_2 b_{22}
\end{array}
\right)=0.
\label{matriu}
\end{equation}
If $\omega \neq 1$, 
a solution with non-vanishing off-diagonal elements is allowed if
$\omega^2=1$, i.e., $\omega=-1$
This leads to
\begin{equation}
A=\left(
\begin{array}{cc}
1 & 0 \\
0 & -1 
\end{array}\right), \;\;\; 
B=\left(
\begin{array}{cc}
0 & b \\
b^* & 0 
\end{array} \right),
\label{unmig}
\end{equation}
which can always be transformed to $\sigma_x$ and $\sigma_y$, by
rotations and adequate normalization.
These are the operators of the example (\ref{tres}). For
spin 1/2 the set of GHZ operators are in this sense unique.

\section{Spin one}
For higher spins the proof proceeds along the same lines. We find one 
case of interest, with $\omega=-1$,
\begin{equation}
A=\left(
\begin{array}{ccc}
1 &  &\\
  &-1& \\
  &  & -1 
\end{array}\right), \;\;\; 
B=\left(
\begin{array}{ccc}
0 & b & c \\
b^* & 0 & 0 \\
c^* & 0 & 0
\end{array} \right).
\label{un}
\end{equation} 
%The eigenstates of $B$ are $0$, $\pm \sqrt{|b|^2+|c|^2}$.
In the basis where $B$ is diagonal $A$ and $B$ read
\begin{equation}
A=-\left(
\begin{array}{ccc}
1 & 0 &0\\
 0 &0 &1 \\
 0 & 1 & 0 
\end{array}\right),\;\;\;
B=\sqrt{|b|^2+|c|^2}\left(
\begin{array}{ccc}
0 & & \\
  & 1 & \\
  &   & -1 
\end{array} \right)
\label{adiag}
\end{equation} 
which proves the assertion in the case of spin one, as a rotation around $x$
brings $B$ into the form $0 \oplus \sigma_y$, while $A$ is left as 
$1 \oplus \sigma_x$, up to normalizations.

\section{Spin 3/2}
For spin 3/2, in addition to cases that reduce straightforwardly 
to those of 
lower spins, we find:
\begin{equation}
A=\left(
\begin{array}{cccc}
1 &  & & \\
  &-1& & \\
  &  &-1 & \\
  &  & &-1 
\end{array}
\right),\;\;
B=\left(
\begin{array}{cccc}
0 & a & b & c \\
a^* & 0 & 0 & 0 \\
b^* & 0 & 0 & 0 \\
c^* & 0 & 0 & 0
\end{array}
\right).
\label{quatre}
\end{equation}
In the basis where $B$ is diagonal $A$ and $B$ read
\begin{equation}
A=-\left(
\begin{array}{cccc}
0 & 1 & 0 & 0 \\
1 & 0 & 0 & 0 \\
0 & 0 & 1 & 0 \\
0 & 0 & 0 & 1
\end{array}
\right),\;\;\;
B=\sqrt{|a|^2+|b|^2+|c|^2} \left(
\begin{array}{cccc}
1 & & & \\
  & -1 & & \\
 & & 0 &  \\
 & & & 0 
\end{array}
\right)
\label{cinc}
\end{equation}
which is again diagonal in two, 2x2, blocks.

The last case corresponds to 
\begin{equation}
A=\left(
\begin{array}{cccc}
1 &  & & \\
  &-1& & \\
  &  & 1 & \\
  &  & &-1 
\end{array}
\right),\;\;
B=\left(
\begin{array}{cccc}
0 & a & 0 & b \\
a^* & 0 & c^* & 0 \\
0 & c & 0 & d \\
b^* & 0 & d^* & 0
\end{array}
\right).
\label{sis}
\end{equation}
The following list of unitary transformations bring these matrices 
to the desired form:

a) With 
\begin{equation}
F=\left(
\begin{array}{cccc}
1 &0 &0 &0 \\
0 &0 &1 &0 \\
0 &1 &0 &0 \\
0 &0 &0 &1 
\end{array}
\right)
\label{flip},
\end{equation}
$F^\dagger=F=F^{-1}$, we find
\begin{equation}
A'=FAF=
\left(
\begin{array}{cc}
I &   \\
  &-I \\
\end{array}
\right),\;\;
B'=FBF=\left(
\begin{array}{cc}
 & {\cal B} \\
{\cal B}^\dagger &
\end{array}
\right),
\label{vuit}
\end{equation}
where
$${\cal B}=\left(
\begin{array}{cc}
a & b \\
c & d 
\end{array}
\right).$$

b) A unitary transformation of the form 
$U=\left(
\begin{array}{cc}
U_1&  \\
   & U_2 
\end{array}
\right)$
leaves $A'$ invariant and allows to diagonalize ${\cal B}$
\begin{equation}
A''=A', \;\;\; B''=UB' U^\dagger=\left(
\begin{array}{cc}
 & U_1 {\cal B} U_2^\dagger \\
(U_1 {\cal B} U_2^\dagger)^\dagger 
\end{array}
\right)=\left(
\begin{array}{cccc}
 & & m & 0 \\
 & & 0 & n \\
m^* & 0 & & \\
0 & n^* & & 
\end{array}
\right).
\label{nou}
\end{equation}
We have used the result that the generic matrix ${\cal B}$ can be brought
to a diagonal form with two unitary matrices $U_1$, $U_2$.

c) Finally, acting with $F$ again,
\begin{equation}
A'''=A, \;\;\; B'''=\left(
\begin{array}{cccc}
0 & m & & \\
m^* & 0 & & \\
 & & 0 & n \\
 & & n^* & 0
\end{array}
\right),
\end{equation}
which completes the proof.

\section{Conclusions}
We conclude that the equation $AB=\omega BA$ is very restrictive on $\omega$
and on the
possible forms of A and B; as the Hilbert space dimension
increases, with increasing spin, all its solutions 
for $\omega \neq 1$ have $\omega=-1$ and
are essentially direct
sums of the two-dimensional $\sigma_x$ and $\sigma_y$. In this
sense there are no solutions that could, in principle, enrich
the possibilities opened by the GHZ theorem.

\vspace*{1cm}
\section{Acknowledgments}
J.S., J.T., R.T. acknowledge the Centro de Ciencias de Benasque for 
hospitality while this work was beeing done. J.S. also acknowledges the 
Department ECM for hospitality and financial support. J.T. and R.T.
acknowledge financial support by CICYT project AEN 98-0431,
CIRIT project 1998 SGR-00026 and CEC project IST-1999-11053.

\end{document}